# Where Chicagoans tweet the most: Semantic analysis of preferential return locations of Twitter users

Aiman Soliman, Junjun Yin, Kiumars Soltani, Anand Padmanabhan, Shaowen Wang

CyberGIS Center for Advanced Digital and Spatial Studies,
University of Illinois at Urbana-Champaign, Urbana, IL 61801
{asoliman, jyn, soltani2, apadmana, shaowen}@illinois.edu

## ABSTRACT

Recent studies on human mobility show that human movements are not random and tend to be clustered. In this connection, the movements of Twitter users captured by geo-located tweets were found to follow similar patterns, where a few geographic locations dominate the tweeting activity of individual users. However, little is known about the semantics (landuse types) and temporal tweeting behavior at those frequently-visited locations. Furthermore, it is generally assumed that the top two visited locations for most of the users are home and work locales (**Hypothesis A**) and people tend to tweet at their top locations during a particular time of the day (**Hypothesis B**). In this paper, we tested these two frequently cited hypotheses by examining the tweeting patterns of more than 164,000 unique Twitter users whom were residents of the city of Chicago during 2014. We extracted landuse attributes for each geo-located tweet from the detailed inventory of the Chicago Metropolitan Agency for Planning. Top-visited locations were identified by clustering semantic enriched tweets using a DBSCAN algorithm. Our results showed that although the top two locations are likely to be residential and occupational/educational, a portion of the users deviated from this case, suggesting that the first hypothesis oversimplify real-world situations. However, our observations indicated that people tweet at specific times and these temporal signatures are dependent on landuse types. We further discuss the implication of confounding variables, such as clustering algorithm parameters and relative accuracy of tweet coordinates, which are critical factors in any experimental design involving Twitter data.

## Categories and Subject Descriptors
J.4 **[Social and Behavioral Sciences]**: Sociology

## General Terms
Measurement, Experimentation, Human Factors

## Keywords
Twitter, Big Data, social media, human mobility, semantic trajectories, urban activity

## 1. INTRODUCTION
### 1.1 Related Work
Understanding human mobility patterns in urban environments is critical for organizing transportation, urban planning and studying the spread of infectious diseases [1-3]. Recently, the availability of new sources of geo-located Big Data has shown promise in studying the complex human mobility patterns [4]. For example, users' trajectories, which were extracted from phone call history [1] revealed that human movements are not random and characterized by a preferential return to few frequently-visited locations [2, 5]. Identifying these frequently-visited locations and their semantics (e.g., home, work) is essential for understanding the origin and destination of each trip and the purpose of human mobility at both the individual and collective levels. Although human movements could be extracted from different data sources (e.g. phone call data, GPS trajectories, survey data), there are advantages of using Twitter data over other Big data sources in studying human mobility patterns. For example, geo-located Twitter records provide a consistent data at regional scales, which is available to public without infringement on individual privacy in case of using phone call records, and the positional accuracy of each geo-locate tweets, approximately 10 meters, provides adequate resolution to resolve the associated landuse type particular in mixed landuse neighbors [5-7].

Analysis of geo-located Twitter data showed the presence of preferential return locations, in which most of the uses' broadcasting activities take place near few geographic locations [6]. Different methods were used to identify preferential return locations for Twitter users or spatial clustering of big data. The simplest method is to impose a fixed grid over the study area and identify the cells with largest number of geo-located tweets [8]. More advanced techniques include bivariate normal density kernel [9], modified mean shift algorithm [10] or applying DBSCAN algorithm to geo-located tweets collected during night (7:00 P.M. to 4:00 A.M.) [11]. Both mean shift and DBSCAN algorithms do not require the number of clusters in advance, which is advantageous because the number of frequently-visited locations could vary from one user to another.

### 1.2 Contribution
Although the literatures provide various examples of detecting preferential return locations of Twitter users, the semantic composition (i.e. landuse type) of these top-visited locations was not examined in detail. For example, the top two visited locations, where users spent most of their time, are usually assumed to be home and work locales without verifications [5, 6, 11]. These assumptions could result in a higher uncertainty in identifying the purpose of individual and collective mobility patterns and understanding intercity dynamics. In this context, we present a better method for extracting and labeling frequently-visited locations of Twitter users. We examined the semantic composition of top-visited locations of residents of

Chicago by matching them to a detailed landuse map, which is to the best of our knowledge, the first attempt to incorporate and analyze the semantics of top tweeted locations. In addition, we introduce a semantic uncertainty index as an objective method for assessing the quality of identified preferential return locations and to account for issues such as GPS accuracy and the geographic spread of geo-located tweets as factors that could impact the quality of spatial clustering. Finally, we demonstrate that the distribution of hourly volume of tweets at these preferential return locations have distinct time signatures that are dependent on the dominant landuse type and discuss its potential application for labeling frequently visited locations.

## 2. DATA AND METHODS

Geo-located Twitter data was collected over North America using Twitter streaming API from January 1st through December 31st, 2014 within a bounding box with lower left and upper right corners' coordinates 41.201577N, -88.707599W, 42.495775N, -87.524535W respectively. Each tweet in the final data set contained a geo-tag and a timestamp indicating local time in CDT, which was corrected for daylight saving calendar. The trajectory of each unique Twitter user *(x, y, time)* was constructed by arranging a user's locations recorded during 2014 in chronological order. We selected the data above the median value, which yielded 164,409 (48.8%) unique users with an average of 64 tweets per user, a maximum of 2683 tweets and minimum of 4 tweets (median value). The landuse inventory for Northeastern Illinois is one of the most detailed and updated landuse maps of Chicago [12]. We retained 23 classes from the original landuse classes that were found to be informative, while merging the remaining excessively detailed classes into seven meaningful classes. Following, we constructed the semantic trajectories by associating each geo-located tweet with one of the thirty landuse types *(x, y, time, landuse)*. We decided to reassign geo-located tweets that fall on the road networks to the nearest landuse polygon since it is unlikely that streets could count as top-visited locations.

We applied the DBSCAN clustering algorithm [13] to each unique user trajectory $T_s$ to identify clusters of semantic-enriched tweets, which should corresponded to the top-visited locations for that user. We defined a search window ($\varepsilon$) of approximately 250 meters (0.0025 degrees) [5] to account for variability of GPS accuracy between devices and the fact that top-visited locations are not infinitesimal points on a map. The minimum number of points to form a cluster was selected to be four to ensure that it is a true location and not merely a coincidence. After applying the DBSCAN clustering algorithm we were able to associate most of the tweets to one of the top-visited clusters, which were ordered in a descending fashion based on the number of associated tweets *(x, y, time, landus_j, cluster rank)*.

The accuracy of any identified top visited location (named here semantic uncertainty) was assessed using the ratio of tweets that belongs to the dominant landuse type to the total number of tweets at each frequently-visited location. We pooled the dominant landuse types for top-visited locations for all unique users residing of Chicago to test the hypothesis that the top two visited locations are home and work locales (Hypothesis A). Furthermore, we tested the hypothesis that users are likely to tweet from their top-visited location at a specific time of the day (Hypothesis B). We tested Hypothesis B by combining timestamps from all tweets and unique users that belong to a certain rank of top-visited locations into a single pool, which was then examined for the existence of significant features.

## 3. RESULTS AND DISCUSSIONS

### 3.1 Semantics of Frequently-Visited Locations

The absolute count of frequently-visited locations decreased exponentially by increasing the rank of top-visited locations (Figure 1). The power distribution signifies that the majority of Chicago residents tweet near one or the top two locations at the most and it is less likely to find users who frequently tweet from a large number of locations. These results are in agreement with the findings of Song et al. [2] which suggested that the top two locations explained more than 60% of the total frequent visitations' pattern. The most significant result is that, although the most likely location for Chicagoans to tweet from is their home (i.e. residential land use type), there are significant cases in which home did not count as the most frequently-visited location. The results summarized in Figure 1 indicate that Hypothesis A is oversimplifying the real world. We also speculate that the percentage of residential landuse types identified at higher ranks exists to complete the cases in which home were not at the top tweeting location. However, a complete understanding of this effect needs a full analysis of the transition matrix between different landuse types.

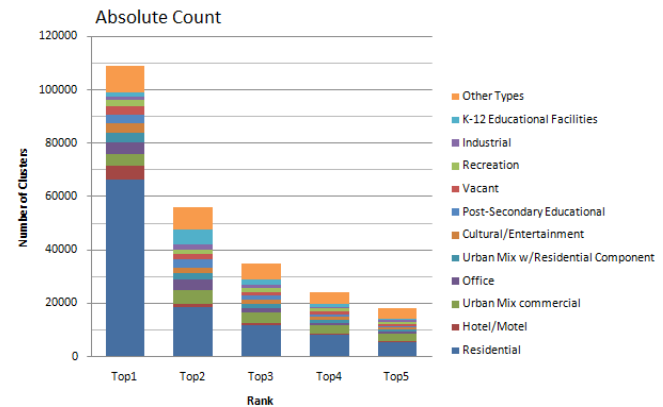

**Figure.1 Semantic composition of top-visited location for all residents of Chicago**

Nevertheless, the likelihood of finding a frequently-visited location with a particular landuse type is still dependent on the rank of frequency of visitation. For example, it is most likely to find dominant landuse type such as schools, post secondary education and office at the second top location and similarly hotels are likely to be found on the top-visited local (Figure 2).

We plotted the distribution of uncertainty ratio for all identified top-visited locations at the top five ranks for all landuse types (Figure 3). The distribution of semantic uncertainty shows that the majority of clusters identified using DBSCAN have a purity ratio that is no less than 65%, as indicated by the lower 25 percentile edge for all five ranks. However, the semantic purity decreased with increasing the rank, which could be attributed to the spread of tweets as the frequency of visitation decrease.

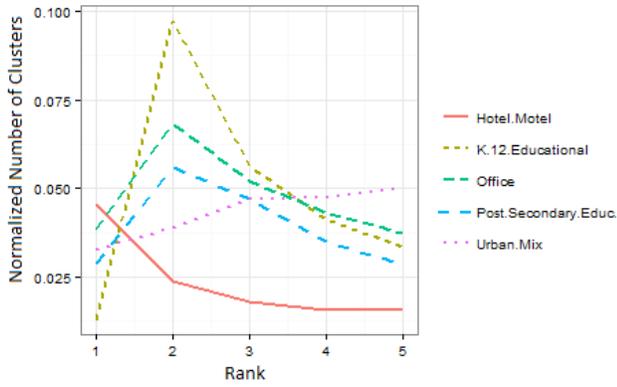

**Figure 2.** Dependency of landuse type on the rank of frequency of visitation, excluding residential landuse for visualization; clusters count was normalized to the total number of clusters in the rank

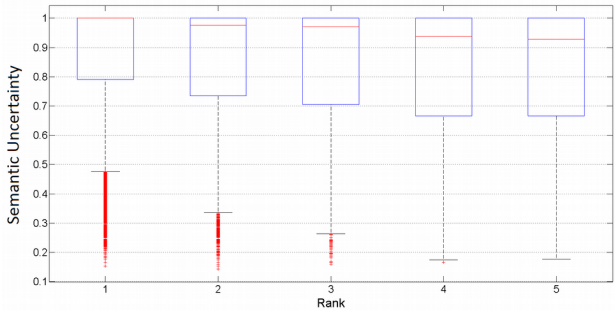

**Figure 3.** Boxplot of semantic uncertainty range for all top-visited locations arranged in a descending order, where rank 1 represents the most visited location; the semantic uncertainty is scaled between 0 and 1. For example 0.7 means that 70 % of the tweets were identified with the dominant landuse type

### 3.2 Sensitivity of DBSCAN Parameters

DBSCAN algorithm requires selecting two parameters, the size of searching window ($\varepsilon$) and the minimum number of points in any cluster. In order to investigate the impact of parameters' sensitivity on our results, we compared our initial selection (a 250 meter search window and a minimum of 4 points) with a more restrictive set of parameters (500 meters and a minimum of 10% of the users' tweets), hereafter named Experiment 2. The results (Table 1) indicated that restricting clustering parameters resulted in a smaller number of clusters for most landuse types, with exception to residential and K12 educational.

Remarkably, the number of hotels and cultural/entertainment clusters dropped significantly for Experiment 2 (Table 1). We speculate that relaxing the clustering parameters in Experiment 1 allowed detection of the behavior of temporary residents and tourists, which is characterized by a single top location (hotel) with fewer tweets and scattered visitation near touristic attraction sites. By restricting the clustering parameters in Experiment 2, the Post-Secondary Educational landuse type became the second most common landuse type found at top-visited location 1 after residential, which is likely because university students spent most of their time on campus. In summary, the results in Table 1 suggest that tuning DBSCAN parameters could allow filtering some generic human movement's patterns.

**Table 1.** The impact of DBSCAN parameters on the percentage of dominant landuse type at Top Location 1; Exp.1 ($\varepsilon$=250 m, minimum of 4 points) and Exp.2 ($\varepsilon$=500 m, minimum of 10% of total user tweets)

| Top Location 1 | Exp. 1 | Exp. 2 |
|---|---|---|
| Residential | 61.1 | 76.1 |
| Hotel | 4.5 | 1 |
| Urban mix Com. | 4.3 | 2.9 |
| Office | 3.8 | 2.5 |
| Urban mix Resi. | 3.3 | 2.8 |
| Cultural | 3.3 | 0.7 |
| Post Sec. Educ. | 2.9 | 2.9 |

### 3.3 Temporal Tweeting Patterns

We examined the hourly patterns of tweeting in order to test if people tend to tweet at their top-visited locations at particular time of the day (Hypothesis B). Figure 4 depicts the hourly distribution of tweets at the most visited location (Top Location 1) classified by different land use types. The observations clearly demonstrate that tweeting intensity at top-visited Location 1 varies considerably between different landuse types. We also noticed that the observed variations of tweeting intensity for different landuse types are tightly coupled with their function. For example, K-12 educational landuse is associated with a peak in the morning and a significant drop after 3:00 P.M. local time, which is consistent with students' daily activity. Similarly, the temporal signature associated with Office landuse type showed a similar diurnal pattern like schools, but with less decline in the afternoon hours as people continue to work after 3 P.M. In contrast, residential, hotel/motel and urban mix with residential units showed all a significant increase of nocturnal activity. Remarkably, the temporal patterns bear clear resemblance to spectral signatures, which are common in remote sensing analysis.

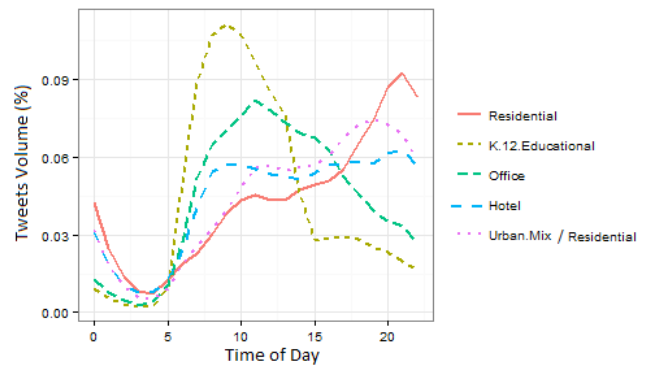

**Figure 4.** Hourly tweeting intensity for most frequently-visited location (Top Location 1) normalized by the total clusters count for each landuse category

Our results depicting the changes of hourly tweeting intensity for different landuse types (Figure 4) compared well with the travel diary survey collected in the city of Chicago, which was measured using survey data collected from 30,000 individuals, refer to Figure 4-a from Jiang et al.[14]. In addition, the comparison between Twitter and survey-based data indicated that tweeting activity might be slightly decoupled from daily activity during certain periods of the day. For example, most of the surveyed individuals indicated the high likelihood to find them at their Home location during early hours of the day (i.e. 1 A.M.to 5 A.M.). However, this period showed a significant drop of tweeting intensity because people are likely to be found at home sleeping. Another example is the apparent peak of tweeting associated with Office landuse type around 12:00 P.M., which could be linked to the possible increase of tweeting activity during lunch breaks. In general, the distinct diurnal variability of tweeting is a very promising approach for distinguishing the semantics of frequently-visited locations of individual users. However, the ultimate success of classifying a cluster to one of the typical landuse types is dependent on the semantic purity of the cluster, as well as the number of tweets involved.

## 4. CONCLUSION AND FUTURE WORK

We demonstrated the potential of using Twitter data in analyzing urban dynamics. In this context, we examined the semantic composition of preferential return locations for Twitter users. We found that home (residential landuse) explained 60-75% of the semantics of all top-visited locations, while offices, schools and universities explained approximately 7, 10 and 6% respectively of the semantics of the second most visited location for all users. Our results suggested that generic assumptions, such as top two visited locations are home and work locales, are oversimplifications of the real world and therefore they are not reliable. However, the hourly volumes of tweets provide a more generic method to label these frequently-visited locations because they are tightly coupled with the landuse types of preferential return locations. The presented case study is a work in progress to build a scalable workflow for the synthesis of geo-located Twitter data with more authoritative data sources. Pinpointing the frequently-visited locations is the first step to analyze the origin-destination of users' trips and to understand the purpose of these trips as well as the function of different neighbor within metropolitan cities based on the collective mobility patterns of its residents.

## 5. ACKNOWLEDGMENTS

This research is based in part upon work supported by the U.S. National Science Foundation under grant numbers: 1047916, 1443080, and 1354329. Insightful comments were received from members of the CyberGIS Center for Advanced Digital and Spatial Studies.